\documentclass[structabstract]{aa}  
\usepackage{graphicx}
\usepackage{txfonts}
\usepackage{natbib}
\usepackage[breaklinks=false]{hyperref}
\newcommand\dif{{\rm d}}

\newcommand\kb{\ensuremath{k_{\rm B}}}
\newcommand\mh{\ensuremath{m_{\rm H}}}
\newcommand\asb{\ensuremath{a_{\rm SB}}}
\newcommand\Ms{{\ensuremath{\mathrm{M}_{\odot}}}}

\newcommand\Mpy{\Ms\,{\rm yr}{\ensuremath{^{-1}}}}

\newcommand\pgaz{{\ensuremath{P_{\rm gas}}}}
\newcommand\prad{{\ensuremath{P_{\rm rad}}}}

\newcommand\srad{{\ensuremath{s_{\rm rad}}}}
\newcommand\rhoc{{\ensuremath{\rho_c}}}
\newcommand\Mcore{{\ensuremath{M_{\rm core}}}}
\newcommand\phicore{{\ensuremath{\varphi_{\rm core}}}}
\newcommand\phisurf{{\ensuremath{\varphi_{\rm surf}}}}
\newcommand\dxi{{\ensuremath{\dif\xi}}}
\newcommand\dphi{{\ensuremath{\dif\varphi}}}

\begin{document}

\title{General-relativistic instability in rapidly accreting supermassive stars in the presence of dark matter}

\author{
L. Haemmerl\'e
}
\authorrunning{Haemmerl\'e}

\institute{D\'epartement d'Astronomie, Universit\'e de Gen\`eve, chemin des Maillettes 51, CH-1290 Versoix, Switzerland}


 
\abstract
{The collapse of supermassive stars (SMSs) via the general-relativistic (GR) instability
would provide a natural explanation to the existence of the most extreme quasars.
The presence of dark matter in SMSs is thought to potentially impact their properties,
in particular their mass at collapse.
Dark matter might be made of weakly interactive massive particles (WIMPs)
that can be captured by the gravitational potential well of SMSs due to interaction with the baryonic gas,
favouring high dark matter densities in the star's core.
The annihilation of WIMPs can provide fuel to support the star before H-burning ignition,
favouring low densities of baryonic gas, long stellar lifetimes and high final masses.}
{Here, we estimate the impact of dark matter on the GR dynamical stability of rapidly accreting SMSs.}
{We add a dark matter term to the relativistic equation of adiabatic pulsations and apply it to hylotropic structures
in order to determine the onset point of the GR instability.
We consider both a homogeneous dark matter background
and density profiles of the form $\propto\exp(-r^2/r_\chi^2)$, typical for the case of WIMPs capture.
The free choice of the central temperature in hylotropic models allows to consider SMSs fuelled by H-burning
as well as by WIMP annihilation.}
{We find that, in principle, the dark matter gravitational field can remove completely the GR instability.
However, for SMSs fuelled by H-burning, the dark matter densities required to stabilise the star against GR
are orders of magnitude above the values that are expected for the dark matter background.
In the case of WIMPs capture, where the required densities can be reached in the centre of the star,
the high centralisation of the dark matter component prevents any effect on the GR instability.
On the other hand, for SMSs fuelled by WIMP annihilation, we find that the low densities of baryonic gas
inhibit the destabilising GR corrections,
which shifts the stability limit by typically an order of magnitude towards higher masses.
As long as central temperatures $\lesssim10^7$ K are maintained by WIMP annihilation,
the GR instability is reached only for stellar masses $>10^6$ \Ms.}
{Dark matter can impact the GR dynamical stability of SMSs
only in the case of energetically significant WIMP annihilation.
The detection of a SMS with mass $>10^6$ \Ms\ in an atomically cooled halo
can be interpreted as an evidence of WIMP annihilation in the star's core.}

 
\maketitle
%

\section{Introduction}
\label{sec-in}

Supermassive black hole (SMBH) formation is currently one of the most challenging open problems in astronomy
(e.g.~\citealt{woods2019,haemmerle2020a}).
At redshifts $z\gtrsim7$, a number of SMBHs with masses $M_\bullet\gtrsim10^9$~\Ms\
were detected in the last decade
(e.g.~\citealt{wu2015,banados2018,yang2020,wang2021a}), which implies average accretion rates above 1~\Mpy.
Such rapid mass growths are in tension with the scenario of an initial seed
provided by the collapse of a $\sim100$~\Ms\ star (e.g.~\citealt{wang2021a}),
and point towards the 'direct collapse' scenario,
i.e. the direct formation of a seed with $M_\bullet\sim10^5-10^7$~\Ms\
from the monolithic collapse of gas at galactic scales (e.g.~\citealt{rees1978}).
This tension is reinforced by the recent discoveries made with the James Webb Space Telescope at higher redshifts,
for instance GN-z11 ($z=11$, $M_\bullet\sim10^6$~\Ms, \citealt{maiolino2023}),
UHZ1 ($z=10.3$, $M_\bullet\sim4\times10^7$~\Ms, \citealt{bogdan2024})
and GHZ 9 ($z\simeq10$, $M_\bullet\sim0.5-1\times10^8$~\Ms, \citealt{kovacs2024}).
Interestingly, the masses of these objects correspond to those of seeds generated by direct collapse,
while no SMBH is found with a mass below those allowed by this scenario \citep{maiolino2023c,andika2024}.

Direct collapse implies the existence of supermassive stars (SMSs) with masses $>10^4$~\Ms\
(e.g.~\citealt{fowler1966,baumgarte1999b,hosokawa2013,shibata2016a,umeda2016,woods2017,woods2020,woods2021a,woods2024,haemmerle2018a,haemmerle2018b,haemmerle2019c,haemmerle2020c,haemmerle2021b,nagele2022,nagele2023a,nagele2023c,nagele2023b,herrington2023,nandal2023})
and their eventual collapse via the general-relativistic (GR) instability \citep{chandrasekhar1964}.
Two channels of direct collapse have been considered in the literature:
the collapse of a primordial atomically cooled halo
(e.g.~\citealt{haiman1997a,omukai2001a,dijkstra2008,bromm2003b,latif2013e,regan2017,chon2018}),
in which a Population~III (Pop~III) SMS forms under accretion rates 0.01 -- 1 \Mpy;
the mergers of massive, gas-rich galaxies \citep{mayer2010,mayer2015,mayer2024},
triggering inflows of 100 -- 1000 \Mpy\ down to sub-parsec scales,
and resulting potentially in the formation of a Population~I (Pop~I) SMS
\citep{haemmerle2019c,haemmerle2021c,zwick2023}.

Without dark matter, the mass of the black hole seed,
given essentially by the final mass of the progenitor SMS at collapse,
belongs to distinct ranges in these two scenarios \citep{haemmerle2020c,haemmerle2021b}:
$10^5\ \Ms\lesssim M\lesssim10^6$ \Ms\ in atomically cooled halos;
$M>10^6$ \Ms\ in galaxy mergers.
On the other hand, the presence of a dark matter background
has been shown to delay the GR instability via its gravitational field, in the case of polytropic models
\citep{mclaughlin1996,bisnovatyi1998,kehrer2024}.
But polytropic SMSs correspond to isentropic, thermally relaxed structures,
which stands in conflict with accretion at rates $\gtrsim0.01$ \Mpy\
\citep{hosokawa2013,umeda2016,woods2017,haemmerle2018a,haemmerle2019c}.
For such accretion rates, the structure is better approximated by 'hylotropes'
\citep{begelman2010,haemmerle2020c,haemmerle2021b},
made of a convective, isentropic core and a radiative envelope.
Moreover, as pointed out by \cite{bisnovatyi1998},
the stabilisation of a polytropic SMS by the gravitational field of a dark matter background
requires unrealisticly high densities of dark matter ($\gtrsim0.001$ g cm$^{-3}$).
Models of adiabatic contraction for star formation indicate ambiant dark matter densities
$\lesssim10^{-10}$ g cm$^{-3}$ in protostars \citep{freese2009}.
On the other hand, if dark matter is made of weakly interactive massive particles (WIMPs),
the densities can be increased by orders of magnitude, thanks to the interactions with the baryonic gas,
that allows to capture the dark matter particles in the gravitational potential well of the star
\citep{spolyar2008,taoso2008,taoso2010,freese2010,kouvaris2011,ellis2022}.
In this case, the captured dark matter component converges to a highly centralised thermal density profile
$\propto\exp(-r^2/r_\chi^2)$ \citep{taoso2008}.
Moreover, WIMPs being Majorana particles, they are their own antiparticles
and can annihilate to convert their mass (in the range 1 GeV -- 10 TeV, typically 100 GeV) into thermal energy.
For efficient enough capture, the energetic input from WIMP annihilation
is found to prevent the contraction of stars before H-burning ignition \citep{spolyar2008,taoso2008,freese2010},
favouring low densities of baryonic gas and low central temperatures.
Models of SMSs fuelled by WIMP annihilation indicate that the central temperature is kept down to $10^6-10^7$ K
for objects of $10^5-10^7$~\Ms,
in contrast with the $\sim10^8$ K given by H-burning.
In this picture, WIMP annihilation limits the accumulation of dark matter in the star's core,
avoiding catastrophic collapse (in contrast with \citealt{kouvaris2011,ellis2022}).
We also notice that, in the case of energetically significant WIMP annihiliation,
the low central temperature shall translate into a reduction of the neutrino emission studied in \cite{kehrer2024}.

In the present work, we study how dark matter impacts the maximum mass of SMSs
consistent with GR dynamical stability.
We apply the post-Newtonian method of \cite{mclaughlin1996} to the full range of hylotropes,
considering both homogeneous and $\propto\exp(-r^2/r_\chi^2)$ density profiles for the dark matter component,
in order to determine the impact of the dark matter gravitational field on the GR stability of SMSs.
We consider the central temperatures given by H-burning,
as well as those imposed by energetically significant WIMP annihilation.
The method is presented in Sect.~\ref{sec-meth};
the results are described in Sect.~\ref{sec-res} and discussed in Sect.~\ref{sec-dis};
we conclude in Sect.~\ref{sec-out}.

\section{Method}
\label{sec-meth}

\subsection{GR instability}

A sufficient condition for GR instability can be obtained from the relativistic equation of adiabatic pulsations
(e.g.~\citealt{chandrasekhar1964,fowler1966,fuller1986,haemmerle2021a,feng2021c}).
This criterion is based on a mathematical proof by \cite{chandrasekhar1964},
with the same GR ingredients used in numerical stellar models.
SMSs are always close to the Newtonian and Eddington limits,
and in this case the criterion for instability reads $\omega^2<0$ with \citep{haemmerle2021b}:
\begin{equation}
\omega^2I=\int\beta P\dif V-\int\left({2GM_r\over rc^2}+{8\over3}{P\over\rho c^2}\right){GM_r\over r}\dif M_r
\label{eq-gr}\end{equation}
where $r$ is the radial coordinate, $M_r$ the mass-coordinate, $P$ the pressure, $\beta$ the ratio of gas to total pressure, $\rho$ the density of mass,
$\dif V$ a volume element, $I$ the moment of inertia, $G$ the gravitational constant and $c$ the speed of light.
The ratio $\beta$ is related to the entropy of radiation \srad\ by
\begin{equation}
\beta={\pgaz\over P}={1\over1+{\prad\over\pgaz}}={1\over1+{\mu\mh\over4\kb}\srad}
\end{equation}
where \kb\ is the Boltzmann constant, \mh\ the mass of a baryon and $\mu$ the mean molecular weight.
And because of the proximity with the Eddington limit, it is well approximated by
\begin{equation}
\beta\simeq{4\kb\over\mu\mh}s^{-1}\ll1
\label{eq-beta}\end{equation}
where $s\simeq\srad$ is the total specific entropy.

In this post-Newtonian approximation, distinctions between relativistic- and rest-mass can be neglected.
Moreover, the criterion~(\ref{eq-gr}) becomes an exact condition \citep{chandrasekhar1964}.
Dynamical simulations in full GR indicate the instability slightly earlier \citep{nagele2022},
but the discrepancies remain in the range of second-order post-Newtonian corrections.

On the other hand, if we account for a dark matter component with density $\rho_\chi\ll\rho$,
interacting with the star via gravitation only without following the stellar pulsations,
a Newtonian pulsation analysis gives \citep{mclaughlin1996}
\begin{equation}
\omega^2I=\int\beta P\dif V+{8\pi G\over3}\int\rho_\chi r^2\dif M_r
\label{eq-dm}\end{equation}
If the dark matter distribution is homogeneous, equation~(\ref{eq-dm}) becomes
\begin{equation}
\omega^2I=\int\beta P\dif V+4\pi G\rho_\chi I
\label{eq-dm0}\end{equation}
We see that dark matter stabilises the star via a positive term in the pulsation equation:
if a Lagrangian layer contracts slightly from its equilibrium position,
the decrease in the enclosed mass corresponding to the dark matter component
will reduce the gravity on this layer and favour the restoring force.
Thus, the assumption that dark matter does not follow the pulsations
corresponds to the case where its stabilising effects are the strongest.
If dark matter follows the pulsations, it behaves like baryonic gas,
and can be neglected in the dynamical stability since $\rho_\chi\ll\rho$.
We emphasise that we do not account for the eventual instability of the dark matter component itself,
in particular we do not consider the possibility of a runaway contraction of dark matter,
in contrast with \cite{kouvaris2011} and \cite{ellis2022}:
our treatment focuses on the GR instability of the baryonic gas.

As long as the GR corrections for the gas remain small, and as long as $\rho_\chi\ll\rho$,
the GR corrections for dark matter are negligible compared to the Newtonian dark matter term and the post-Newtonian term for gas \citep{mclaughlin1996,bisnovatyi1998}.
In this case, the combined effect of GR corrections and dark matter is well captured by
\begin{eqnarray}
\omega^2I&=&\int\beta P\dif V+{8\pi G\over3}\int\rho_\chi r^2\dif M_r	\nonumber\\
&&\qquad\qquad\qquad	-\int\left({2GM_r\over rc^2}+{8\over3}{P\over\rho c^2}\right){GM_r\over r}\dif M_r
\label{eq-grdm-1}\end{eqnarray}
or, for homogeneous dark matter distribution:
\begin{equation}
\omega^2I=\int\beta P\dif V+4\pi G\rho_\chi I-\int\left({2GM_r\over rc^2}+{8\over3}{P\over\rho c^2}\right){GM_r\over r}\dif M_r
\label{eq-grdm}\end{equation}
This equation can be re-expressed in dimensionless form by defining the following dimensionless functions:
\begin{equation}
\xi=\alpha r,									\quad
\rho=\rho_c\theta^3,								\quad
M_r={4\over\sqrt{\pi}}\left({K\over G}\right)^{3/2}\varphi,	\quad
P=P_c\psi.
\label{eq-adim}\end{equation}
The constants $K$ and $\alpha$, that give respectively the mass- and length-scales, are defined from the central pressure and density by
\begin{equation}
P_c=K\rhoc^{4/3}	\qquad
\alpha^2={\pi G\rho_c^2\over P_c}={\pi G\rho_c^{2/3}\over K}
\label{eq-alpha}\end{equation}
We define the following parameter for the GR corrections:
\begin{equation}
\sigma={P_c\over\rhoc c^2}
\end{equation}
With these definitions, equation~(\ref{eq-grdm}) reads
\begin{eqnarray}
{\omega^2\over\omega_0^2}\cdot\left({8\over3}\int\xi^2\dphi\right)
=\beta_c\cdot\left(\int{\beta\over\beta_c}\psi\xi^2\dxi\right)
+{\rho_\chi\over\rhoc}\cdot\left({8\over3}\int\xi^2\dphi\right)	\nonumber\\
-\sigma\cdot\left(32\int\left({\varphi\over\xi}+{\psi\over3\theta^3}\right){\varphi\over\xi}\dphi\right)
\label{eq-gradim}\end{eqnarray}
with $\omega_0^2:=4\pi G\rhoc$, where $\beta_c$ is the value of $\beta$ at the centre of the star.
Each term of equation~(\ref{eq-gradim}) is made of an integral that involves only the dimensionless functions (\ref{eq-adim}),
to which we add $\beta/\beta_c$ for $\beta$:
\begin{eqnarray}
I_0&:=&{8\over3}\int\xi^2\dphi	\\
I_+&:=&\int{\beta\over\beta_c}\psi\xi^2\dxi	\\
I_-&:=&32\int\left({\varphi\over\xi}+{\psi\over3\theta^3}\right){\varphi\over\xi}\dphi
\end{eqnarray}
Each integral is scaled by a dimensionless parameter:
$\beta_c$ expresses the departures from the Eddington limit, $\sigma$ the departures from the Newtonian limit, and $\rho_\chi/\rhoc$ the presence of dark matter.
The limit of dynamical stability is reached when the left-hand side, and so the right-hand side, vanish:
\begin{eqnarray}
\beta_cI_++{\rho_\chi\over\rhoc}I_0-\sigma I_-=0
\label{eq-grbeta}\end{eqnarray}
By their definitions, $\beta_c$ and $\sigma$ are related by the central temperature and the chemical composition:
\begin{equation}
\beta_c\sigma={\kb T_c\over\mu\mh c^2}
\end{equation}
We can use this relation to eliminate $\sigma$ in (\ref{eq-grbeta}):
\begin{eqnarray}
\beta_c^2+{\rho_\chi\over\rhoc}{I_0\over I_+}\beta_c-{\kb T_c\over\mu\mh c^2}{I_-\over I_+}=0
\label{eq-grbeta2}\end{eqnarray}
For a given dimensionless structure $(\xi,\theta,\phi,\psi,\beta/\beta_c)$, a given central temperature and a given chemical composition,
equation~(\ref{eq-grbeta2}) can be solved for $\beta_c$ as a function of the ratio $\rho_\chi/\rhoc$:
\begin{eqnarray}
\beta_c=\sqrt{{\kb T_c\over\mu\mh c^2}{I_-\over I_+}+\left({\rho_\chi\over\rhoc}{I_0\over2 I_+}\right)^2}-{\rho_\chi\over\rhoc}{I_0\over2 I_+}
\label{eq-grbeta2sol}\end{eqnarray}
Alternatively, it can be solved as a function of the density $\rho_\chi$ itself
by eliminating $\rhoc$ with the definition of $\beta_c$:
\begin{equation}
\beta_c={{\kb\over\mu\mh}\rhoc T_c\over P_c}\simeq{{\kb\over\mu\mh}\rhoc T_c\over{1\over3}\asb T_c^4}	\quad\Longrightarrow\quad
\rhoc={\mu\mh\over\kb}{\asb\over3}T_c^3\beta_c
\label{eq-rhoc-betac}\end{equation}
\begin{eqnarray}
\beta_c^2={\kb T_c\over\mu\mh c^2}\left({I_-\over I_+}-{\rho_\chi c^2\over{1\over3}\asb T_c^4}{I_0\over I_+}\right)
\label{eq-grbeta1}\end{eqnarray}
In this case, the equation for $\beta_c$ at the limit of stability does not necessarily have a real root.
Indeed, if
\begin{eqnarray}
{\rho_\chi c^2\over{1\over3}\asb T_c^4}>{I_-\over I_0}
\label{eq-rhocrit}\end{eqnarray}
the right-hand side of (\ref{eq-grbeta1}) is negative and so the condition for stability is always satisfied:
\begin{eqnarray}
\beta_c^2>0>{\kb T_c\over\mu\mh c^2}\left({I_-\over I_+}-{\rho_\chi c^2\over{1\over3}\asb T_c^4}{I_0\over I_+}\right)
\label{eq-betarhocrit}\end{eqnarray}
In other words, if there is enough dark matter to satisfy (\ref{eq-rhocrit}),
the GR instability is completely removed for a given $T_c$ and a given dimensionless structure ($I_0$ and $I_-$).
This effect does not appear when we consider a fixed ratio $\rho_\chi/\rho_c$
because, since $\rho_c$ decreases as the stellar mass increases,
a fixed ratio $\rho_\chi/\rho_c$ corresponds to a decreasing value of $\rho_\chi$,
weakening the stabilising effect of dark matter.
But as long as $\rho_\chi$, $T_c$, $I_0$ and $I_-$ are maintained constant,
the instability can be reached only if the gas density decreases down to dark matter density,
for the destabilising post-Newtonian corrections to the dark matter gravitational field to become significant.
Otherwise, the final mass of the SMS is set by fuel exhaustion instead of GR instability.

\subsection{Hylotropes}

The internal distribution of entropy (\ref{eq-beta}) in rapidly accreting SMSs is well approximated by
\begin{equation}
\beta=\left\{\begin{array}{cc}	\beta_c\qquad\qquad				&	{\rm if}\ M_r<\Mcore	\\
						\beta_c\left({M_r\over\Mcore}\right)^{-1/2}	&	{\rm if}\ M_r>\Mcore	\end{array}\right.
\label{eq-bhylo}\end{equation}
The dimensionless ratio $\Mcore/M$ determines the relative distribution of entropy over the mass,
and once it is fixed, all the dimensionless functions (\ref{eq-adim}) are determined.
These dimensionless structures are called 'hylotropes' \citep{begelman2010}, and are thus uniquely classified by the dimensionless parameter $\Mcore/M$.
In particular, for each ratio $\Mcore/M$, we know the values of all the dimensionless integrals in equation~(\ref{eq-grbeta}).
Once all the dimensionless functions are defined, there remain only two free parameters to determine the structure, plus one for dark matter.
For the present purpose, it is convenient to choose the central temperature $T_c$.
If we fix the last free parameter by imposing the critical condition~(\ref{eq-grbeta}), we obtain the properties of SMSs at the limit of stability,
in particular the maximum masses consistent with stability:
\begin{eqnarray}
M&=&{4\over\sqrt{\pi}}\left({K\over G}\right)^{3/2}\phisurf		\label{eq-Mcrit}\\
\Mcore&=&{4\over\sqrt{\pi}}\left({K\over G}\right)^{3/2}\phicore	\label{eq-Mcorecrit}
\end{eqnarray}
We notice that the mass-scale is fully determined by $K$, and for a mixture of ideal gas and radiation,
this constant is fully determined by $\beta_c$ (and the chemical composition):
\begin{equation}
K={P_c\over\rhoc^{4/3}}=\left({3\over\asb}\right)^{1/3}\left({\kb\over\mu\mh}\right)^{4/3}\left({1-\beta_c\over\beta_c^4}\right)^{1/3}
\label{eq-K}\end{equation}

\subsection{WIMPs capture}

In the case of WIMPs capture,
the dark matter component is distributed according to a thermal distribution \citep{taoso2008}:
\begin{equation}
\rho_\chi=\rho_{\chi,c}e^{-{r^2\over r_\chi^2}}		\qquad
r_\chi=\sqrt{3\kb T_c\over2\pi G\rho_c m_\chi}
\label{eq-rchi}\end{equation}
where $m_\chi$ is the mass of the dark matter particles.
The characteristic radius (\ref{eq-rchi}) can be expressed in terms of the length-scale parameter $\alpha^{-1}$ (equations \ref{eq-adim}-\ref{eq-alpha}):
\begin{equation}
r_\chi={\xi_\chi\over\alpha}	\qquad	\xi_\chi=\sqrt{{3\over2}{\mu\mh\over m_\chi}\beta_c}
\end{equation}
\begin{equation}\Longrightarrow\qquad
\rho_\chi=\rho_{\chi,c}e^{-{\xi^2\over\xi_\chi^2}}
\end{equation}
We see that the dimensionless characteristic radius $\xi_\chi$ is proportional to the geometric mean of two ratios:
$\beta_c=\pgaz/P$ and the ratio $\mu\mh/m_\chi$ of the average mass of the free particles of gas to the mass of the dark matter particles.
WIMPS models give typical values $m_\chi\sim100$ GeV,
that is two orders of magnitude larger than the proton mass \mh.
Thus, for $\mu$ of the order of unity and $\beta_c$ of the order of a percent, we have $r_\chi\sim0.01\alpha^{-1}$.
The length-scale $\alpha^{-1}$ corresponds for instance to the ratio of the core radius to the dimensionless core radius $\xi_{\rm core}$.
This last quantity is always in the range 1 -- 7 for all the hylotropes, so that $\alpha^{-1}$ gives the order of magnitude of the core radius,
and the characteristic radius (\ref{eq-rchi}) is always of the order of a percent of the core radius.

\section{Results}
\label{sec-res}

\subsection{Homogeneous dark matter distribution}

\begin{figure}\begin{center}
\includegraphics[width=.5\textwidth]{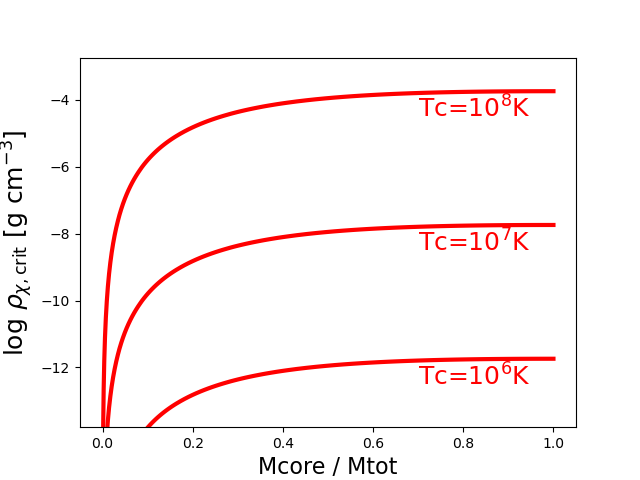}
\caption{Dark matter density threshold above which the GR instability is completely removed
for any stellar mass (equation~\ref{eq-rhocrit}), as a function of the mass fraction of the convective core,
for the indicated central temperatures and assuming homogeneous dark matter density.}
\label{fig-dmcrit}
\end{center}\end{figure}

The density threshold above which dark matter removes completely the GR instability (equation~\ref{eq-rhocrit})
is plotted on figure~\ref{fig-dmcrit} as a function of $\Mcore/M$ for different $T_c$,
assuming homogeneous dark matter density.
A change in $T_c$ corresponds to a simple rescaling of the curve, i.e. a constant shift in logarithmic scales.
For each $T_c$, the function is monotonously growing, and for any given $\rho_\chi$
there is a ratio $\Mcore/M$ below which the GR instability is completely removed.
Moreover, the fact that the function has a maximum, reached in the polytropic limit $\Mcore\to M$,
implies that, for each $T_c$, there is a threshold in $\rho_\chi$ above which
the GR instability is completely removed independently of $\Mcore/M$.
For the typical $T_c$ of H-burning ($\sim10^8$~K), the threshold is $\sim10^{-4}$ g cm$^{-3}$.
More precisely, for Pop III SMSs fuelled by H-burning ($T_c\sim2\times10^8$~K),
the threshold is $\sim3\times10^{-3}$ g cm$^{-3}$,
while in the Pop I case ($T_c=8-9\times10^7$ K) it is $\sim10^{-4}$ g cm$^{-3}$.
In both cases, it corresponds to $\sim10^{-3}\rho_c$.
We emphasise that, thanks to the thermostatic effect of H-burning,
$T_c$ remains nearly constant all along the H-burning phase,
so that dark matter above this density threshold will prevent the GR instability until fuel exhaustion.
For the typical $T_c$ of WIMP annihilation ($\sim10^6-10^7$~K), the threshold is lower:
$\sim10^{-12}-10^{-8}$ g cm$^{-3}$.

\begin{figure}\begin{center}
\includegraphics[width=.5\textwidth]{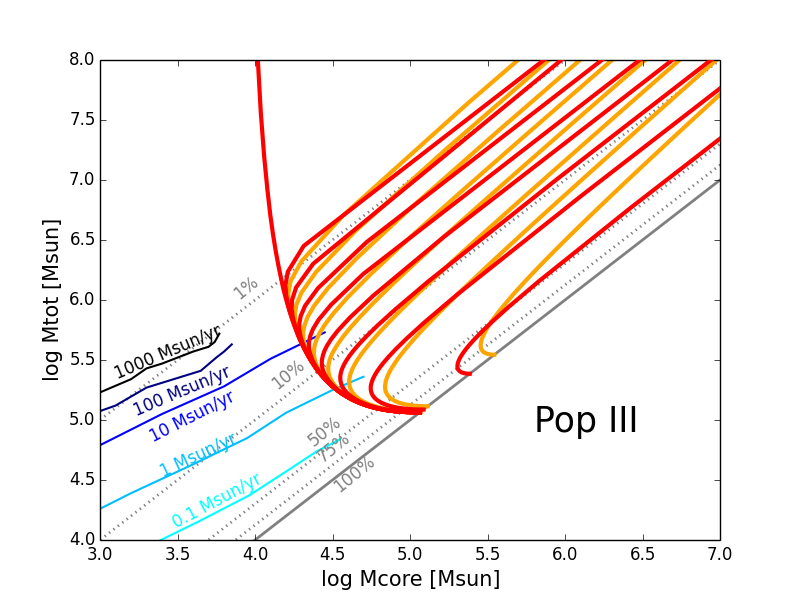}
\caption{Limits of dynamical stability in the $(\Mcore,M)$-diagram for Pop III hylotropic SMSs
($\mu=0.6$, $T_c=1.8\times10^8$~K)
in the presence of a homogeneous dark matter distribution with various densities $\rho_\chi$.
The red curves correspond to $\rho_\chi=0$ and to $\log\rho_\chi\ {\rm[cgs]}=-10,-9,...,-3$ (from left to right);
the orange curves to $\log\rho_\chi/\rhoc=-10,-9,...,-3$ (from left to right).
The black-blue-cyan lines are GENEC tracks for the indicated accretion rates
\citep{haemmerle2018a,haemmerle2019c}.
The grey lines correspond to constant mass fractions of the convective core.}
\label{fig-mmcore}
\end{center}\end{figure}

The stability limits in the $(\Mcore,M)$-diagram are shown in figure~\ref{fig-mmcore}
for a homogeneous dark matter distribution, for $\mu=0.6$ and $T_c=1.8\times10^8$ K,
typical values for Pop~III SMSs in the beginning of the H-burning phase.
For the dark matter parameter, we considered both a constant $\rho_\chi/\rhoc$,
with values $\log\rho_\chi/\rhoc=-10,-9,...,-3$ (orange lines),
and a constant $\rho_\chi$, with values $\rho_\chi=0$ and $\log\rho_\chi\ {\rm[cgs]}=-10,-9,...,-3$ (red lines).
The intersections of the curves correspond to the models where $\rhoc=1$ g cm$^{-3}$.
We notice that the lines of constant $\Mcore/M$ (grey lines)
represent the asymptotes of the stability limits with constant $\rho_\chi$ (red curves).
It reflects the fact that, as noted above, for each $T_c$ and for each $\rho_\chi$,
there is a ratio $\Mcore/M$ below which the GR instability is completely removed (figure~\ref{fig-dmcrit}).

\begin{figure}\begin{center}
\includegraphics[width=.5\textwidth]{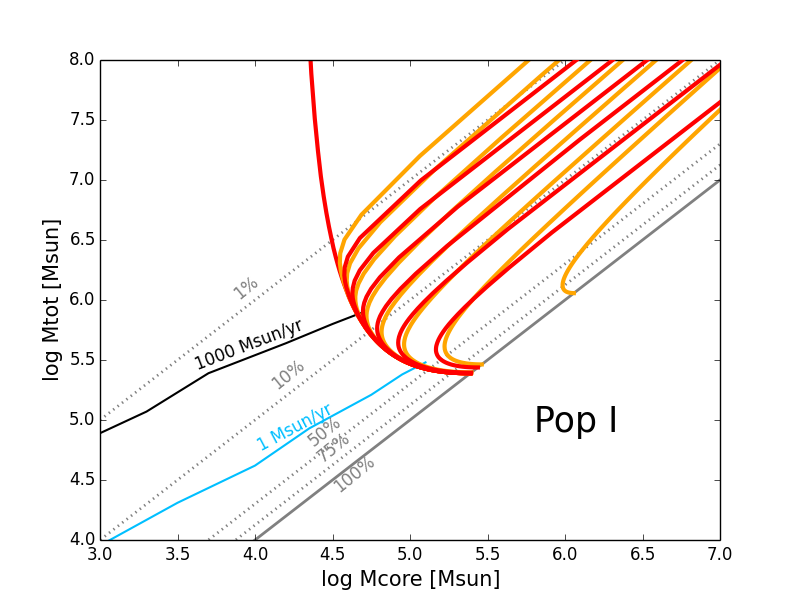}
\caption{Same as figure~\ref{fig-mmcore} for the Pop I case ($\mu=0.6$, $T_c=8.7\times10^7$ K).
From left to right, the red curves correspond to $\rho_\chi=0$ and to $\log\rho_\chi\ {\rm[cgs]}=-10,-9,...,-5$ only,
since the GR instability is completely removed for $\log\rho_\chi\ {\rm[cgs]}\geq-4$.}
\label{fig-mmsol}
\end{center}\end{figure}

The limits are shown in figure~\ref{fig-mmsol} for $\mu=0.6$ and $T_c=8.7\times10^7$ K,
relevent for Pop~I SMSs in the beginning of the H-burning phase.
The orange lines correspond to the same constant values of the ratio $\log\rho_\chi/\rhoc=-10,-9,...,-3$
as in the Pop~III case,
and the red lines to constant values of $\rho_\chi=0$ and $\log\rho_\chi\ {\rm[cgs]}=-10,-9,...$ up to $-5$ only,
because in the Pop~I case larger densities satisfy condition (\ref{eq-rhocrit}),
so that no limit can be shown (as seen in figure~\ref{fig-dmcrit}).
For a given stellar mass, the central density of SMSs is lower in the Pop~I case than in the Pop~III case,
because of an earlier start of H-burning \citep{haemmerle2019c}.
While in the Pop~III case it is typically 1 g cm$^{-3}$ near the instability,
in the Pop~I case it is $\sim0.1$ g cm$^{-3}$,
so that the red curve for $\log\rho_\chi\ {\rm[cgs]}=-n$ intersects the orange one for $\log\rho_\chi/\rhoc=-n-1$.
As a consequence, the intersection of the orange and red curves in figure \ref{fig-mmsol}
corresponds to a central density of 0.1 g cm$^{-3}$.

\begin{figure}\begin{center}
\includegraphics[width=.5\textwidth]{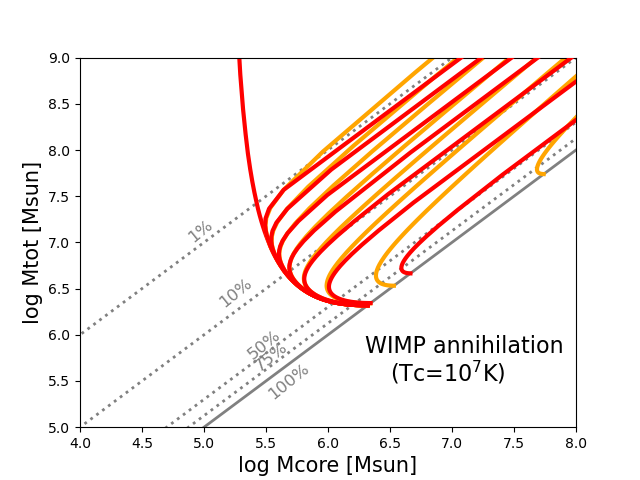}
\caption{Same as figures~\ref{fig-mmcore}-\ref{fig-mmsol} for $\mu=0.6$ and $T_c=10^7$ K,
typical for the case of energetically significant WIMP annihilation.
From left to right, the red curves correspond to $\rho_\chi=0$ and to $\log\rho_\chi\ {\rm[cgs]}=-14,-13,...,-8$ only,
since the GR instability is completely removed for $\log\rho_\chi\ {\rm[cgs]}>-8$.}
\label{fig-mm1e7}
\end{center}\end{figure}

The same limits are shown in figure~\ref{fig-mm1e7} for $\mu=0.6$ and $T_c=10^7$ K,
typical values for SMSs fuelled by WIMP annihilation \citep{freese2010}.
In this case, the red lines correspond to constant values of $\rho_\chi=0$ and $\log\rho_\chi\ {\rm[cgs]}=-14,-9,...$
up to $-8$ only, since for larger densities the GR instability is completely removed (figure~\ref{fig-dmcrit}).

By comparing figures \ref{fig-mmcore}-\ref{fig-mmsol}-\ref{fig-mm1e7},
we see that a decrease in $T_c$ translates into a shift of the stability limit towards higher masses.
While for $T_c\sim10^8$ K the GR instability can be reached at $M\sim10^5$ \Ms\
in the polytropic limit ($\Mcore\to M$),
for $T_c\sim10^7$ K it requires masses $M\gtrsim2\times10^6$ \Ms.
This effect reflects the fact that lower temperatures correspond to lower densities of the baryonic gas,
and thus to smaller GR destabilising corrections.
Moreover, because of the lower density of baryonic gas, the dark matter density required to modify the stability limit
is lower when $T_c$ is lower.
For instance, for $\Mcore\sim10\% M$, while the limit for Pop~III SMSs fuelled by H-burning is modified by dark matter
only for $\rho_\chi\gtrsim10^{-6}$ g cm$^{-3}$ (figure~\ref{fig-mmcore}),
the limit for SMSs fuelled by WIMP annihilation is modified
as soon as $\rho_\chi\gtrsim10^{-10}$ g cm$^{-3}$ (figure~\ref{fig-mm1e7}).

\subsection{Centralised dark matter distribution}
\label{sec-capture}

For the highly centralised distributions (\ref{eq-rchi}), we find that the mass-limit in the $(\Mcore,M)$-diagram
remains identical to the case without dark matter.
These distributions are well approximated by a $\delta$-distribution, which does not impact the stability of the star
(as it can be seen in equations~\ref{eq-dm}-\ref{eq-grdm-1}),
since the enclosed mass remains identical in case of contraction or expansion of a Lagrangian layer.
The stability limit starts to be shifted only for $\xi_\chi\gtrsim10$, that is for $m_\chi\lesssim10^{-4}$ GeV,
orders of magnitude below the mass range of WIMPs models.

\section{Discussion}
\label{sec-dis}

\subsection{Comparison with previous works}

For polytropic SMS models, assuming homogeneous dark matter densities,
\cite{mclaughlin1996} found that the stability limit
start to be significantly shifted towards larger masses as soon as $\rho_\chi\gtrsim10^{-3}\rhoc$,
which corresponds to $\rho_\chi\gtrsim10^{-3}$ g cm$^{-3}$ for Pop III SMSs
and to $\rho_\chi\gtrsim10^{-4}$ g cm$^{-3}$ in the Pop I case.
This threshold is visible figures~\ref{fig-mmcore} and \ref{fig-mmsol} in the polytropic limit $\Mcore\to M$.
But we see that, when $\Mcore\lesssim M/2$, a smaller amount of dark matter is sufficient:
for $\rho_\chi\sim10^{-4}\rhoc$, the maximum mass is increased by a factor of a few
as soon as $\Mcore\sim50\%M$,
and by nearly an order of magnitude when $\Mcore\sim20\%M$;
for $\rho_\chi\sim10^{-5}\rhoc$, it is increased by about an order of magnitude
if the core remains $\Mcore\sim10\%M$.

\subsection{Final mass of SMSs in direct collapse}

In the main chanel of direct collapse,
that is the collapse of a primordial, atomically cooled halo without fragmentation into a central SMBH,
the direct progenitor of the SMBH is a Pop III SMS accreting at rates 0.01 -- 1 \Mpy\
(e.g.~\citealt{latif2013e,chon2018}).
In this case, evolutionary models of SMSs fuelled by H-burning
give $T_c\simeq2\times10^8$ K and $\Mcore\gtrsim10\%M$ \citep{haemmerle2018a,haemmerle2019c},
so that a homogeneous dark matter distribution of $10^{-4}-10^{-3}$ g cm$^{-3}$
is required to increase the final mass by a factor of a few (figure~\ref{fig-mmcore}).
The dark matter density must exceed 0.003 g cm$^{-3}$
for the GR instability to be completely removed until fuel exhaustion (figure~\ref{fig-dmcrit}).
Since the H-burning time of SMSs is around a Myr (e.g.~\citealt{umeda2016,woods2017,woods2020}),
we can estimate the maximum mass of SMSs to be $\sim10^6$ \Ms\ in this case.

In the case of galaxy merger driven direct collapse, the direct progenitor of the SMBH
is a Pop I SMS accreting at rates potentially up to 100 -- 1000 \Mpy\ (e.g.~\citealt{mayer2024}),
and this scenario requires a SMS's mass $\gtrsim10^6$ \Ms\ at collapse (e.g.~\citealt{montero2012}).
Figure~\ref{fig-mmsol} shows that masses in excess of $10^6$ \Ms\ are accessible in this case
for a dark matter background $\gtrsim10^{-5}$ g cm$^{-3}$,
if accretion rates $\sim1000$ \Mpy\ can be maintained on the star.
For $\sim10^{-6}$ g cm$^{-3}$, the maximum mass can be increased by a factor of a few.
The GR instability is completely removed during the H-burning phase
only if the dark matter density is $\gtrsim10^{-4}$ g cm$^{-3}$ (figure~\ref{fig-dmcrit}).
And because of the huge accretion rates accessible in galaxy mergers,
we can expect masses up to $10^8-10^9$ \Ms\ to be accreted in the H-burning time.

For the cases where dark matter removes completely the GR instability in the approximation used here,
the instability shall be reached only if the destabilising post-Newtonian corrections
to the dark matter gravitational field become significant.
It occurs once $\rho_c\sim\rho_\chi$, as $\rho_c$ shall decrease for increasing stellar mass $M$.
For given $T_c$, $\rho_c$ scales with $\beta_c$ (equation~\ref{eq-rhoc-betac}),
which scales with $M^{-1/2}$ (equations~\ref{eq-Mcrit}-\ref{eq-K}).
We see that, for $\rho_\chi\sim10^{-3}$ g cm$^{-3}$, the evolution from $\rho_c\sim0.1-1$ g cm$^{-3}$
to $\rho_c\sim\rho_\chi$ requires a decrease in $\rho_c$ by 2 -- 3 orders of magnitude,
that is an increase in $M$ by 4 -- 6 orders of magnitude.
One can expect fuel exhaustion to occur earlier.

However, the densities expected for the dark matter background a the centre of mini-halos
are $\lesssim10^{-10}$ g cm$^{-3}$ \citep{freese2009}.
Figures~\ref{fig-mmcore}-\ref{fig-mmsol} show that, for such densities,
the stability limit is shifted only for $\Mcore\lesssim1\%M$ for the $T_c$ of H-burning.
But stellar models indicate that $\Mcore\gtrsim5\%M$ for the relevant masses
\citep{haemmerle2018a,haemmerle2019c}.
We see that the stabilisation of a H-burning SMS by the gravitational field of a dark matter background
requires unrealisticly high densities,
and WIMPs capture does not change the picture (Sect.~\ref{sec-capture}).
By comparison, rotation appears as a much more efficient stabilising agent, in agreement with \cite{bisnovatyi1998}.
Indeed, for SMSs accreting angular momentum at the $\Omega\Gamma$-limit, hylotropic models indicate that,
in atomically cooled halos, the final mass is increased by a factor of a few by the effect of rotation,
while in galaxy mergers it is increased by orders of magnitude \citep{haemmerle2021b}.
With baryonic gas and rotation only, the final mass of SMSs is found to belong to distinct ranges:
$10^5\ \Ms<M<10^6\ \Ms$ for atomically cooled halos, and $10^6\ \Ms<M<10^9\ \Ms$ for galaxy mergers.

On the other hand, in the case of WIMP annihilation,
$T_c\sim10^7$ K are found for the relevant masses \citep{freese2010}.
In this case, a homogeneous dark matter background of $\sim10^{-10}$ g cm$^{-3}$
is sufficient to impact the stability limit via its gravitational field if $\Mcore\sim10\%M$ (figure \ref{fig-mm1e7}).
Moreover, independently of the effect of the dark matter gravitational field,
the low temperatures imposed by the energetic effect of WIMP annihilation
lead to a shift of the stability limit towards higher masses.
As long as central temperatures $\lesssim10^7$ K are maintained,
the GR instability is reached only for stellar masses $\gtrsim2\times10^6$ \Ms\ (figure \ref{fig-mm1e7}).
Thus, with WIMP annihilation, SMSs of $M>10^6$ \Ms\ can remain stable
in atomically cooled halos as well as in galaxy mergers, even without any rotation.
The detection of a SMS with such a high mass in an atomically cooled halo
would thus be a sign of energetically significant WIMP annihilation in the core of the star.

\section{Conclusion}
\label{sec-out}

In the present work, we studied the impact of dark matter on the GR instability in SMSs
via the effect of the dark matter gravitational field,
as well as the effect of WIMP annihilation on the thermal state of the star.
We considered hylotropic structures with different ratio of the convective, isentropic core's mass \Mcore\
to the total stellar mass $M$.

Via its gravitational field only, dark matter can in principle remove completely the GR instability in SMSs
until fuel exhaustion (figure~\ref{fig-dmcrit}),
and increase their final masses by orders of magnitude (figures~\ref{fig-mmcore}-\ref{fig-mmsol}-\ref{fig-mm1e7}).
The smaller is the ratio $\Mcore/M$, the stronger is the impact of dark matter on the stability limit.
However, for the expected densities of the dark matter background and for realistic stellar structures,
the effect of the dark matter gravitational field remains negligible.
The higher densities reached in the case of WIMPs capture do not change the picture
because of the high centralisation of the captured particles (Sect.~\ref{sec-capture}).

On the other hand, for the low central temperatures imposed by energetically significant WIMP annihilation,
the stability limit is shifted by an order of magnitude towards higher masses,
compared to the case of SMSs fuelled by H-burning.
While for the central temperatures of H-burning the GR instability can be reached at $M\sim10^5$ \Ms,
for the central temperatures given by WIMP annihilation it requires masses $M\gtrsim2\times10^6$ \Ms.

Thus, we conclude that, in realistic conditions, the gravitational field of dark matter
does not impact the stability limit of SMSs,
but WIMP annihilation can postpone the GR instability and increase the final mass by an order of magnitude.
Overall, we can estimate that, with rotation and energetically significant WIMP annihilation,
the final masses of SMSs formed in atomically cooled halos range in the interval $10^6\ \Ms<M<10^7\ \Ms$.
The detection of a SMS with mass $>10^6$ \Ms\ in an atomically cooled halo
can be interpreted as an evidence of energetically significant WIMP annihilation in the star's core.

\begin{acknowledgements}
LH has received funding from the European Research Council (ERC) under the European Union's Horizon 2020 research and innovation programme
(grant agreement No 833925, project STAREX).
\end{acknowledgements}

\bibliographystyle{aa}
\bibliography{bib}

\end{document}